\documentclass[prd,twocolumn,amsmath,amssymb,nofootinbib,showpacs]{revtex4-1}
\usepackage{graphicx}
\usepackage{dcolumn}
\usepackage{bm}
\usepackage{color}
\usepackage{aas_macros}
\allowdisplaybreaks[1]
\newcommand{\matlabel}[1]{\underline{\underline{#1}}}
\begin{document}
\title{Constraining the photon-axion coupling constant with magnetic white dwarfs}
\author{Ramandeep Gill}
  \email{rsgill@phas.ubc.ca}
\author{Jeremy S. Heyl}
  \email{heyl@phas.ubc.ca}
\affiliation{
Department of Physics and Astronomy, University of British Columbia\\
6224 Agricultural Road, Vancouver, BC V6T 1Z1, Canada
}
\date{\today}
\begin{abstract}
The light pseudoscalar particle, dubbed the axion, 
borne out of the Peccei-Quinn solution to the strong CP 
problem in QCD remains elusive. One avenue of inferring 
its existence is through its coupling to electromagnetic 
radiation. So far, laboratory experiments have dedicated all 
efforts to detect the axion in the mass range $10^{-6} < m_a < 10^{-3}$ 
eV with a photon-axion coupling strength $g_{a\gamma\gamma} 
< 10^{-10}\mbox{ GeV}^{-1}$, where the limits are derived 
from astrophysical considerations. In this study, we present 
a novel way of constraining $g_{a\gamma\gamma}$ by looking 
at the level of linear polarization in the radiation emerging 
from magnetic white dwarfs (mWDs). We find that photon-axion oscillations 
in WD magnetospheres can enhance the degree of linear 
polarization. Observing that most mWDs show only 5\% linear 
polarization, we derive upper limits on $g_{a\gamma\gamma}$ 
for different axion masses.
\end{abstract}
\keywords{}
\pacs{14.80.Va, 97.20.Rp}
\maketitle
\section{Introduction}
Quantum chromodynamics (QCD) has emerged as a phenomenologically accurate theory 
that describes strong interactions 
among the six quark flavors that are bound into two families of hadrons, namely mesons and 
baryons. From experiments, we understand that strong interactions 
enjoy C (charge conjugation), P (parity), T (time reversal) discrete symmetries of nature. Therefore, 
QCD must also obey such symmetries, both separately and any combinations formed thereof \citep{KimCarosi2010}. 
However, CP symmetry is broken in QCD due to the presence of the following term in the 
QCD Lagrangian \citep{Wilczek1978}
\begin{equation}
\mathfrak{L}_{\text{int}}=\left(\frac{\theta g^2}{32\pi^2}\right)\text{tr}\ 
G^{\mu\nu}_a\tilde{G}_{a\mu\nu}
\end{equation}
where $\theta$ is a periodic parameter, $g$ is the QCD coupling constant, $G_{\mu\nu}$ is the 
color field strength tensor, and $\tilde{G}_{\mu\nu}$ is its dual. The value of the $\theta$-parameter 
is not set theoretically, but it can be measured from 
the electric dipole moment of the neutron ($d_n$), for which many theoretical estimates exist but 
we only quote one, $\|d_n\|\sim 2.7\times 10^{-16}\bar{\theta}e$ cm \citep{Baluni1979}. Here $\bar{\theta}
=\theta+\text{arg det }m_q$, where $m_q$ is the quark mass matrix. The latest experimental estimate of 
$\|d_n\| < 2.9\times10^{-26}e$ cm \citep{Bakeretal2006} constrains $\|\bar{\theta}\| \lesssim 10^{-11}$ 
\citep{KimCarosi2010}. This inexplicably small value of $\bar{\theta}$ gave rise to 
the \emph{strong CP problem}. One of the solutions, also the most favoured, to this problem 
was envisioned by Peccei \& Quinn \citep{PecceiQuinn1977}, whereby the $\bar{\theta}$ parameter is driven precisely to 
zero under a global chiral symmetry, later named $U(1)_{\text{PQ}}$. The pseudo-Nambu-Goldstone 
boson that results upon the spontaneous breakdown of this symmetry was dubbed the axion 
\citep{Wilczek1978,Weinberg1978}. Not unlike the Higgs boson, the axion has proven to be extremely 
difficult to observe as it couples only weakly to ordinary matter and radiation.

Despite several attempts to experimentally observe the axion, it remains elusive to this day. 
Nevertheless, the experimental efforts have not gone in vain, but have been able to place serious 
constraints on the coupling strength of the axion to photons $g_{a\gamma
\gamma} < 10^{-10}\text{ GeV}^{-1}$. Stringent constraints have been placed on the mass of the axion 
$10^{-6} \lesssim m_a \lesssim 10^{-3}$ eV with the lower limit arising 
from cosmology \citep{Preskilletal1983} and the upper limit\footnote{Due to large uncertainties in the axion mass derived for the DFSZ 
model \citep{DineFischlerSrednicki1981,Zhitnitskii1980} from SN 1987A observations 
($0.004 \lesssim m_a \lesssim 0.012$ eV), a more relaxed upper limit is 
$m_a \lesssim 0.01$ eV \citep{Raffelt2004}} from the neutrino flux recorded 
for SN 1987A, which placed strong limits on the cooling flux through other channels namely, 
right-handed neutrinos or axions \citep{Raffelt2004}. If $g_{a\gamma\gamma} > 10^{-10}\text{ GeV}^{-1}$, 
the production of axions through the Primakoff process will significantly alter the core He burning timescales of 
post main sequence stars, a possibility excluded by the ratio of horizontal branch stars in 
globular clusters \citep{Raffelt1996}. Several reviews on the properties of axions have been forthcoming in the past 
decade, for example see \citep{Raffelt1999,Asztalosetal2006,KimCarosi2010}, to which we point the 
reader for a more detailed and comprehensive exposition.

Still, there is no denying the fact that none of the laboratory experiments conducted 
thus far have been able to secure a positive detection of this mysterious particle. 
The 
detection of very weakly coupled particles demands extremely sensitive laboratory experiments. 
So far, only a handful of experiments, namely the Cern Axion Solar Telescope (CAST) 
\citep{Andriamonjeetal2007}, Axion Dark Matter Experiment (ADMX) \citep{Asztalosetal2004,Asztalosetal2010}, 
and Rochester-Brookhaven-Fermilab 
(RBF) collaboration \citep{Panfilisetal1987,Wuenschetal1989} have 
been able to surpass the astrophysically derived limits on $g_{a\gamma\gamma}$ in the above 
quoted axion mass range. Yet, the sensitivity envelope needs to be pushed even further by a 
few orders of magitude to be able to draw any definitive conclusions about the existence of 
the axion. Plans are afoot to modifiy the existing experiments and devise new ones to 
improve upon current limits on $g_{a\gamma\gamma}$ (see Section IV).

Unlike the laboratory experiments, the odds are in favour for detecting axions 
in astrophysical systems. This optimism stems from the fact that the 
axion to photon conversion probability scales with large magnetic field strengths and 
longer coherence lengths \citep{Cheloucheetal2009}, such that $P_{a\rightarrow\gamma}
\propto g^2B^2L^2$, where $L$ is the length over which both the photon and axion fields 
are in phase. Thus, there is a very good chance of finding the axion in 
strongly magnetized compact objects, namely magnetic white dwarfs (mWDs) and neutron 
stars (NSs). The possibility in the latter case has been expounded by many (see for 
example \citep{RaffeltStodolsky1988,LaiHeyl2006,Cheloucheetal2009,PshirkovPopov2009}; also see 
\citep{Jimenezetal2011} where constraints on $g_{a\gamma\gamma}$ are derived 
from the dimming of radiation by photon-axion conversion in astrophysical sources), however, the 
case of the mWDs has not been investigated in great detail and warrants further 
study. 
\subsection{Magnetic white dwarfs}
After the discovery of the first mWD by Kemp \citep{Kempetal1970}, the number 
of white dwarfs with magnetic fields ranging from a few kG to $10^3$ MG has grown to about 
170. The size of this subpopulation is only 3\% of the total population of known WDs 
comprising of 5447 objects\footnote{G.P. McCook and E.M. Sion, web version of the Villanova 
White Dwarf Catalog, http://www.astronomy.villanova.edu/WDCatalog/index.html}. The main channel for identifying magnetism 
in WDs is through Zeeman spectropolarimetry, which not only allows one to discern the 
strength of the field but also the direction of the field lines, and also through 
cyclotron spectroscopy (see for e.g. \citep{WickramasingheFerrario2000} for a review 
on isolated and binary mWDs). Nevertheless, reconstruction of the field topology has 
proven to be very difficult, mainly due to its highly non-dipolar structure. Over the 
last decade Zeeman tomography of mWDs has enjoyed some success in elucidating the 
underlying field structure. This technique is based on calculating a database of 
model spectra, where different field geometries comprising of single/multiple dipole, 
and higher multipoles, that may also be off-centered and misaligned with the rotational axis, are 
considered. Then a least-squares fit using the pre-calculated synthetic spectra is performed 
through a highly optimized algorithm on the phase-resolved Zeeman spectra to obtain the 
complex field structures \citep{Euchneretal2002}. The 
generality of the models not only allows greater flexibility but also renders a closer 
fit to the actual field geometry of the source for a given rotational phase.

The presence of even a small degree of circular polarization in the spectrum of a WD is a 
strong indicator of the object possessing a magnetic field upwards of $10^6$ G 
\citep{Kemp1970}. The degree of circular polarization typically reaches up to $\sim 5\%$, 
and sometimes beyond that in a few selective objects, near absorption features and also 
in the continuum. Continuum circular polarization stems from the magnetic circular 
dichroism of the atmosphere, 
where the left and right circularly polarized waves propagating through a magnetized 
medium encounter unequal opacities \citep{Angel1977}. A relatively higher degree of 
circular polarization also appears near the red and blue shifted wings of the Zeeman 
split absorption lines ($\sigma_+$ and $\sigma_-$ components, \citep{WickramasingheFerrario2000}).

On the other hand, most observations of mWDs indicate that the linear polarization 
component never exceeds that of the circular one, and the spectrum remains dominantly 
circularly polarized until field strengths $\geq 10^8$ G are reached \citep{Angel1977}. 
In a magneto-active plasma, the plane of linear polarization undergoes many Faraday 
rotations, an effect that arises due to the magnetic birefringence of the medium, so 
that on average the degree of linear polarization of the emergent radiation is much reduced 
\citep{SazonovChernomordik1975}.

The very fact that linear polarization is of the order of a few percent ($\sim 5\%$) in the 
continuum spectra of most mWDs can be exploited to draw meaningful conclusions on the extent of axion 
interaction with photons traversing the magnetized plasma of mWDs. We explain how this can be 
implemented in the next section.
\subsection{Plan of this study}
The purpose of this study is to conduct a survey of the $m_a-g_{a\gamma\gamma}$ parameter 
space by modelling photon-axion oscillations in the magnetosphere of a mWD. To this end, 
we model the field structure of a strongly magnetized WD PG 1015+014, for which high 
resolution optical spectropolarimetric observations are available \citep{Euchneretal2006}. In the 
same article, the authors also conduct a phase-resolved Zeeman tomographic analysis and derive 
a best-fit model of the magnetic field topology. Despite fitting the spectrum with a range of 
field geometries, they were only able to pin down the field geometry for a single rotational phase 
by fitting it with a superposition of three off-centered and non-aligned dipoles of unequal surface 
field strengths (see Table \ref{tab:pg1015} for model parameters). To model the effect of photon-axion oscillation 
in the magnetosphere on the emergent polarization, we propagate an unpolarized photon of a given 
energy from the photosphere through the encompassing magnetosphere, that has been populated by 
a diffuse, cold ionized H gas. The emergent intensity and polarization is then averaged over the 
whole surface of the star. Finally, we compare the degree of polarization from our model simulation 
to what is observed in mWDs with field strengths in excess of a few $10^6$ G, 
for example PG 1015+014, and draw conclusions on the strength of the coupling constant for 
a given axion mass.
\begin{table}
\caption{\textit{Top}: Magnetic field geometry adopted from the spectropolarimetric analysis by \citep{Euchneretal2006} 
of the the mWD PG 1015+014. The model comprises of three off-centered and non-aligned dipoles 
$D_1,D_2,D_3$ with unequal surface field strengths $B_s$, polar ($\theta_B$) and azimuthal ($\phi_B$) angles 
of the magnetic field axes. The center positions of the dipoles relative to the center of the star are 
given by ($a_x,a_y,a_z$). \textit{Bottom}: Collection of parameters used in the study.}
\begin{center}
\begin{tabular}{l r r r}
\multicolumn{4}{c}{\textbf{Model Parameters}} \\
\hline
\hline
& $D_1$ & $D_2$ & $D_3$ \\
\hline
$B_s$ (MG) & -40 & 92 & -38 \\
$\theta_B({}^\circ)$ & 44 & 63 & 63 \\
$\phi_B({}^\circ)$ & 339 & 276 & 134 \\
$a_x(R_\star)$ & 0.04 & -0.012 & 0.27 \\
$a_y(R_\star)$ & 0.35 & -0.136 & 0.080 \\
$a_z(R_\star)$ & 0.33 & -0.28 & 0.21 \\
\hline
$R_\star$ & \multicolumn{3}{r}{$7\times10^8$ cm} \\
$\theta_k$ & \multicolumn{3}{r}{$23^\circ$} \\
$Y_e$ & \multicolumn{3}{r}{1} \\ 
$B_Q$ & \multicolumn{3}{r}{$4.413\times10^{13}$ G} \\
$T$ & \multicolumn{3}{r}{$10^4$ K} \\
$g_\star$ & \multicolumn{3}{r}{$10^8\text{ cm s}^{-2}$} \\
$\rho_0$ & \multicolumn{3}{r}{$10^{-10}\text{ g cm}^{-3}$} \\
$\rho_\infty$ & \multicolumn{3}{r}{$10^{-20}\text{ g cm}^{-3}$} \\
\hline
\end{tabular} 
\end{center}
\label{tab:pg1015}
\end{table}

In the following Section, we formulate the key equations describing the interaction of the axion 
with photons, geometry of the aggregate magnetic field, and structure of the plasma permeating the 
magnetosphere. The lack of understanding of the density profile of the magnetospheric plasma 
introduces some level of inaccuracy in any treatment of mWDs. We take the simplest approach and 
describe the plasma density by the barometric law for an isothermal atmosphere. In Section III we present 
the main results of the study along with a comparison to some of the results obtained from 
lab experiments (see Fig. \ref{fig:exclusion}). A discussion of the results is provided in 
Section IV.
\section{Model equations}
The interaction of the axion field with an external electromagnetic field is given by the 
following Lagrangian density \citep{RaffeltStodolsky1988}, in natural units where $\hbar=c=1$,
\begin{eqnarray}
\mathfrak{L}=&-\frac{1}{4}F_{\mu\nu}F^{\mu\nu}+\frac{1}{2}(\partial_\mu a
\partial^\mu a - m_a^2 a^2)-\frac{1}{4}g_{a\gamma\gamma}F_{\mu\nu}
\tilde{F}^{\mu\nu}a& \nonumber \\
& +\frac{\alpha^2}{90m_e^4}\left[\left(F_{\mu\nu}F^{\mu\nu}
\right)^2+\frac{7}{4}\left(F_{\mu\nu}\tilde{F}^{\mu\nu}\right)^2\right]&
\label{eq:lagrangian}
\end{eqnarray}
where the first term describes the external electromagnetic field, with $F_{\mu\nu}$ 
as the antisymmetric electromagnetic field strength tensor and $\tilde{F}^{\mu\nu}=
\frac{1}{2}\varepsilon^{\mu\nu\rho\sigma}F_{\rho\sigma}$ as its dual. The second term 
is simply the Klein-Gordon equation for the axion field $a$ where $m_a$ represents its 
mass. The next term is the interaction Lagrangian density, which upon simplification, 
using the given definitions, yields $\mathfrak{L}_{\text{int}}=g_{a\gamma\gamma}
a \mathbf{E}\cdot\mathbf{B}$, where $g_{a\gamma\gamma}$ is the photon-axion coupling 
strength, $\mathbf{E}$ is the polarization vector of the 
photon field, and $\mathbf{B}$ is the external magnetic field. Quantum corrections to 
the classical electromagnetic field, due to the constant creation and annihilation of 
electron-positron pairs in vacuum, are given in the last term of \eqref{eq:lagrangian} by 
the Euler-Heisenberg effective Lagrangian, causing the vacuum to be birefringent.
\subsection{Off-centered non-aligned three dipole model}
\begin{figure}
\includegraphics[width=0.48\textwidth]{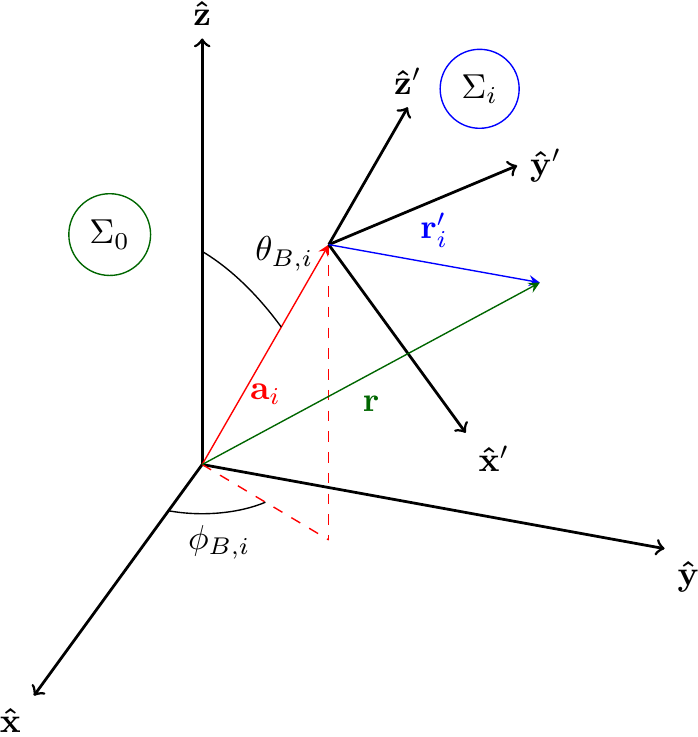}
\caption{This illustrates the coordinate system used to obtain the aggregate magnetic field. 
Here $\Sigma_0$ represents the coordinate system centered on the star with the 
$\mathbf{\hat{z}}$-axis aligned with the rotational axis. $\Sigma_i$ represents the coordinate system in which the 
different dipole magnetic field equations are written. This system is misaligned with $\Sigma_0$ 
by a polar angle $\theta_{B,i}$ and an azimuthal angle $\phi_{B,i}$, and then it is displaced from 
the center by the vector $\mathbf{a}_i$. For clarity we have chosen the vector $\mathbf{a}_i$ to 
lie along the $\mathbf{\hat{z}}'$-axis (color online).}
\label{fig:coord}
\end{figure}
We start by writing down the aggregate magnetic field $\mathbf{B}_0$ in the coordinate 
system $\Sigma_0$ where the $\mathbf{\hat{z}}$-axis coincides with the rotational axis of the star 
\begin{equation}
\mathbf{B}_0=\sum_{i=1}^3R_i^T\mathbf{B}_i'
\end{equation}
The three off-centered dipole fields $\mathbf{B}_i'$ are first rotated before they are added 
together by operating on each of them with a rotation matrix $R^T=R_{z,i}^TR_{y,i}^T$, 
where the superscript $T$ indicates the transpose, and the rotation matrices are 
given as follows
\begin{eqnarray}
R_{z,i}=&\left(\begin{array}{ccc}
\cos \phi_{B,i} & \sin \phi_{B,i} & 0 \\
-\sin \phi_{B,i} & \cos \phi_{B,i} & 0 \\
0 & 0 & 1 \\
\end{array}\right)& \\
R_{y,i}=&\left(\begin{array}{ccc}
\cos\theta_{B,i} & 0 & -\sin\theta_{B,i} \\
0 & 1 & 0 \\
\sin\theta_{B,i} & 0 & \cos\theta_{B,i} \\
\end{array}\right)&
\label{eq:Ry}
\end{eqnarray}
Here the polar and azimuthal angles $\theta_{B,i}$ and $\phi_{B,i}$, respectively, are defined 
with respect to the axis of rotation. In cartesian coordinates, the dipole fields are 
expressed as
\begin{equation}
\mathbf{B}'_i=\frac{B_{s,i}R_\star^3}{2r'^5_i}(3x'_iz'_i\mathbf{\hat{x}'}
+3y'_iz'_i\mathbf{\hat{y}'}+\{3{z'_i}^2-{r'_i}^2\}\mathbf{\hat{z}'})
\end{equation}
where the fields are shifted from the coordinate center, such that $\mathbf{r}_i'=
\mathbf{r}-\mathbf{a}_i$. In the above equation, $\mathbf{B}_{s,i}$ is the surface field 
strength of the $i^{\text{th}}$ dipole field component, $R_\star\simeq7\times10^8$ cm is the radius of the WD, and $r_i'$ is the 
magnitude of the radial vector in the coordinate system $\Sigma_i$. The profile of the aggregate field $B_0$ as a function of distance 
is shown in Fig. \ref{fig:los}.
\subsection{Fully ionized pure H atmosphere}
The presence of a magnetic field necessarily introduces anisotropy in the plasma dielectric tensor 
$\matlabel{\varepsilon}_p$. In the case of a nonuniform field, none of the dielectric tensor components 
vanish, as compared to the homogeneous case. Below we write all the dielectric components, 
which one can easily derive from Maxwell's equations, for completeness.
\begin{align}
\matlabel{\varepsilon}_p & = \left(\begin{array}{ccc}
\varepsilon_{11} & \varepsilon_{12} & \varepsilon_{13} \\
\varepsilon_{21} & \varepsilon_{22} & \varepsilon_{23} \\
\varepsilon_{31} & \varepsilon_{32} & \varepsilon_{33} \\
\end{array}\right) & \\
\varepsilon_{11} & = 1-\sum_{s=e,p}\hat{\omega}_{p,s}^2\left[\frac{1-\hat{\omega}_{c,s}^2
\hat{B}_{0x}^2}{1-\hat{\omega}_{c,s}^2}\right] \nonumber & \\
& \approx 1-\hat{\omega}_{p,e}^2\left[\frac{1-\hat{\omega}_{c,e}^2(1+\hat{\omega}_{c,p}^2)
\hat{B}_{0x}^2}{(1-\hat{\omega}_{c,e}^2)(1-\hat{\omega}_{c,p}^2)}\right] & \\
\varepsilon_{12} & \approx \frac{\hat{\omega}_{c,e}\hat{\omega}_{p,e}^2}{(1-\hat{\omega}_{c,e}^2
)}(i\hat{B}_{0z}+\hat{\omega}_{c,e}\hat{B}_{0x}\hat{B}_{0y}) & \\
\varepsilon_{13} & \approx -\frac{\hat{\omega}_{c,e}\hat{\omega}_{p,e}^2}{(1-\hat{\omega}_{c,e}^2)}
(i\hat{B}_{0y}-\hat{\omega}_{c,e}\hat{B}_{0x}\hat{B}_{0z}) & \\
\varepsilon_{21} & \approx -\frac{\hat{\omega}_{c,e}\hat{\omega}_{p,e}^2}{(1-\hat{\omega}_{c,e}^2)}
(i\hat{B}_{0z}-\hat{\omega}_{c,e}\hat{B}_{0x}\hat{B}_{0y}) & \\
\varepsilon_{22} & \approx 1-\hat{\omega}_{p,e}^2\left[\frac{1-\hat{\omega}_{c,e}^2(1+\hat{\omega}
_{c,p}^2)\hat{B}_{0y}^2}{(1-\hat{\omega}_{c,e}^2)(1-\hat{\omega}_{c,p}^2)}\right] & \\
\varepsilon_{23} & \approx \frac{\hat{\omega}_{c,e}\hat{\omega}_{p,e}^2}{(1-\hat{\omega}_{c,e}^2)}
(i\hat{B}_{0x}+\hat{\omega}_{c,e}\hat{B}_{0y}\hat{B}_{0z}) & \\ 
\varepsilon_{31} & \approx \frac{\hat{\omega}_{c,e}\hat{\omega}_{p,e}^2}{(1-\hat{\omega}_{c,e}^2)}
(i\hat{B}_{0y}+\hat{\omega}_{c,e}\hat{B}_{0x}\hat{B}_{0z}) & \\ 
\varepsilon_{32} & \approx -\frac{\hat{\omega}_{c,e}\hat{\omega}_{p,e}^2}{(1-\hat{\omega}_{c,e}^2)}
(i\hat{B}_{0x}-\hat{\omega}_{c,e}\hat{B}_{0y}\hat{B}_{0z}) & \\
\varepsilon_{22} & \approx 1-\hat{\omega}_{p,e}^2\left[\frac{1-\hat{\omega}_{c,e}^2(1+\hat{\omega}
_{c,p}^2)\hat{B}_{0z}^2}{(1-\hat{\omega}_{c,e}^2)(1-\hat{\omega}_{c,p}^2)}\right] &
\end{align}
In the above set of equations, $\hat{\omega}_{c,s}=q_sB_0/\omega m_s c$ is the normalized cyclotron 
frequency for species $s=(e,p)$, where $e$ and $p$ signify electrons and protons; $\hat{\omega}_{p,s} 
= \sqrt{4\pi n_s/m_s\omega^2}$ is the normalized plasma frequency, where $n_p = n_e = Y_e\rho/m_p$ 
are the electron and proton number densities, $Y_e$ is the electron fraction, and $\rho$ is the 
proton mass density of the plasma; the normalized magnetic field components are defined as 
$\hat{B}_{0,i=x,y,z} = B_{0,i}/B_0$.
\subsubsection{Vacuum corrections}
Due to the polarizability of the vacuum in strong magnetic fields, the plasma dielectric 
tensor $\matlabel{\varepsilon}_p$, and the inverse permeability tensor $\matlabel{\mu}^{-1}$ 
are modified \citep{MeszarosVentura1979,LaiHo2003b}, such that $\matlabel{\varepsilon}_{p+v} = 
\matlabel{\tilde{\varepsilon}} = \matlabel{\varepsilon}_p + \Delta\matlabel{\varepsilon}_v$ 
and $\matlabel{\mu}^{-1}_{p+v} = \matlabel{\tilde{\mu}}^{-1} = \matlabel{I} + 
\Delta\matlabel{\mu}^{-1}_v$, where
\begin{eqnarray}
& \Delta\matlabel{\varepsilon}_v = (a_v-1)\matlabel{I} + q_v\hat{\mathbf{B}}_0\hat{\mathbf{B}}_0 & \\
& \Delta\matlabel{\mu}^{-1}_v = (a_v-1)\matlabel{I} + m_v\hat{\mathbf{B}}_0\hat{\mathbf{B}}_0 & \\
& a_v = 1-2\delta_v\ \ \ q_v = 7\delta_v\ \ \ m_v=-4\delta_v\ \ \ \delta_v = 
\frac{\alpha}{45\pi}\left(\frac{B_0}{B_Q}\right)^2 & \nonumber
\end{eqnarray}
and $B_Q = 4.413\times10^{13}$ G is the quantum critical field for which the separation in energy 
between Landau levels of the electron exceeds its rest mass.
\subsubsection{Plasma density profile}
That many mWDs are surrounded by hot coronae has been suggested by many to explain the polarized 
flux of those WDs that show comparable degree of linear and circular polarization 
\citep{SazonovChernomordik1975,InghamBrecherWasserman1976,ZheleznyakovSerber1994}. The thermal electrons in the hot 
tenuous plasma with temperature $T\sim10^{6-8}$ K radiate at the cyclotron frequency that falls in the 
optical wavelength for field strengths of $B\sim10^8$ G. This radiation appears 
to be polarized both linearly and circularly, depending on the orientation of the line of sight to the 
magnetic field, and traverses the corona without any apsorption. Furthermore, slightly polarized 
radiation emanating from the photosphere, with very low degree of linear polarization due 
to Faraday rotation, gets added to that generated in the corona, as a result increasing the amount 
of flux that is polarized linearly. Several hot isolated WDs, with effective temperatures in 
excess of $\simeq25,000$ K, emitting X-rays were detected by ROSAT \citep{Flemingetal1996}, however 
all cases were linked to subphotospheric thermal emission \citep{Musielaketal2003}. Although the 
non-detection of any coronal emission may indicate the absence of a hot tenuous corona, it is not 
at all unreasonable to suggest the presence of a tenuous cold plasma of fully ionized H. In this 
study, we envisage that the mWDs are encompassed by cold isothermal electron-proton coronae with 
the following barometric density profile,
\begin{equation}
\rho(r) = \rho_0\exp\left(-\frac{r-R_\star}{H_\rho}\right) + \rho_\infty
\end{equation}
where $\rho_0$ is the density near the surface of the star, $\rho_\infty$ is the density 
that remains far away from the star as the strength of the magnetic field becomes significantly 
weaker than that at the surface, and $H_\rho = 2k_BT/m_pg_\star \simeq 1.65\times10^4$ cm is the density 
scale height with an effective temperature $T\simeq10^4$ K and surface gravity 
$\log g_\star(\text{cm/s}^2) = 8$. There is no clear agreement on the surface plasma 
density with $10^{-11} \lesssim \rho_0 \lesssim 10^{-6}\text{ g cm}^{-3}$. Here, we 
assume that the plasma is sufficiently tenuous with $\rho_0 = 10^{-10}\text{ g cm}^{-3}$ 
and $\rho_\infty = 10^{-20}\text{ g cm}^{-3}$.
\subsection{Axion-photon mode evolution in an inhomogeneous magnetized plasma} 
We are interested in knowing the evolution of the axion field and the polarization vector 
as the radiation propagates out from the surface of the star, traversing the region with an 
inhomogeneous plasma density and magnetic field. Here we follow the discussion given in 
\citep{RaffeltStodolsky1988,LaiHeyl2006}, and derive the photon field mode evolution from the 
EM wave equation
\begin{equation}
\nabla\times(\matlabel{\tilde{\mu}}^{-1}\cdot\nabla\times\mathbf{E}) = \frac{\omega^2}{c^2}
\matlabel{\tilde{\varepsilon}}\cdot\mathbf{E}
\end{equation}
Next, we assume the ansatz $\mathbf{E}=\tilde{\mathbf{E}}\exp(ikz)$ where the wave is propagating 
along the rotational axis of the star, which in this case is also the line of sight direction, 
and the wavenumber $k=\omega/c$. Plugging this ansatz into 
the wave equation, and ignoring second order derivatives, we find
\begin{equation}
\frac{d}{dz}
\left(\begin{array}{c}
\tilde{E}_x \\
\tilde{E}_y
\end{array}\right) = \left(\begin{array}{cc}
\chi_{11} & \chi_{12} \\
\chi_{21} & \chi_{22} 
\end{array}\right)\left(\begin{array}{c}
\tilde{E}_x \\
\tilde{E}_y
\end{array}\right)
\label{eq:modeevol}
\end{equation}
where the matrix elements are given below
\begin{eqnarray}
\chi_{11} & = & \Upsilon^{-1}_3\left[k^2\tilde{\varepsilon}_{11}-\Upsilon_4-
\left(1-\frac{\Upsilon_1^2}{\Upsilon_3\Upsilon_5}\right)^{-1} \Upsilon_1
\Upsilon_5^{-1} \right. \nonumber \\
& & \left. \times\left(k^2\tilde{\varepsilon}_{21}-\Upsilon_2-
\frac{\Upsilon_1}{\Upsilon_3}\{k^2\tilde{\varepsilon}_{11}-\Upsilon_4\}
\right)\right] \\
\chi_{12} & = & \Upsilon^{-1}_3\left[k^2\tilde{\varepsilon}_{12}-\Upsilon_2-
\left(1-\frac{\Upsilon_1^2}{\Upsilon_3\Upsilon_5}\right)^{-1} \Upsilon_1
\Upsilon_5^{-1} \right. \nonumber \\
& & \left. \times\left(k^2\tilde{\varepsilon}_{22}-\Upsilon_6-
\frac{\Upsilon_1}{\Upsilon_3}\{k^2\tilde{\varepsilon}_{12}-\Upsilon_2\}
\right)\right] \\
\chi_{21} & = & \left(1-\frac{\Upsilon_1^2}{\Upsilon_3\Upsilon_5}\right)^{-1} 
\Upsilon_5^{-1} \nonumber \\ 
& & \times \left(k^2\tilde{\varepsilon}_{21}-\Upsilon_2-
\frac{\Upsilon_1}{\Upsilon_3}\{k^2\tilde{\varepsilon}_{11}-\Upsilon_4\}
\right) \\
\chi_{22} & = & \left(1-\frac{\Upsilon_1^2}{\Upsilon_3\Upsilon_5}\right)^{-1} 
\Upsilon_5^{-1} \nonumber \\ 
& & \times \left(k^2\tilde{\varepsilon}_{22}-\Upsilon_6-
\frac{\Upsilon_1}{\Upsilon_3}\{k^2\tilde{\varepsilon}_{12}-\Upsilon_2\}
\right) \\
\Upsilon_1 & = & \frac{d}{dz}(m_v\hat{B}_x\hat{B}_y)+i2km_v\hat{B}_x\hat{B}_y \\
\Upsilon_2 & = & ik\frac{d}{dz}(m_v\hat{B}_x\hat{B}_y)-k^2m_v\hat{B}_x\hat{B}_y \\
\Upsilon_3 & = & -\frac{d}{dz}(a_v+m_v\hat{B}_y^2)+i2k(a_v+m_v\hat{B}_y^2) \\
\Upsilon_4 & = & -ik\frac{d}{dz}(a_v+m_v\hat{B}_y^2)+k^2(a_v+m_v\hat{B}_y^2) \\
\Upsilon_5 & = & -\frac{d}{dz}(a_v+m_v\hat{B}_x^2)+i2k(a_v+m_v\hat{B}_x^2) \\
\Upsilon_6 & = & -ik\frac{d}{dz}(a_v+m_v\hat{B}_x^2)+k^2(a_v+m_v\hat{B}_x^2)
\end{eqnarray}
\subsubsection{Line of sight geometry}
\begin{figure}[t!]
\includegraphics[width=0.5\textwidth]{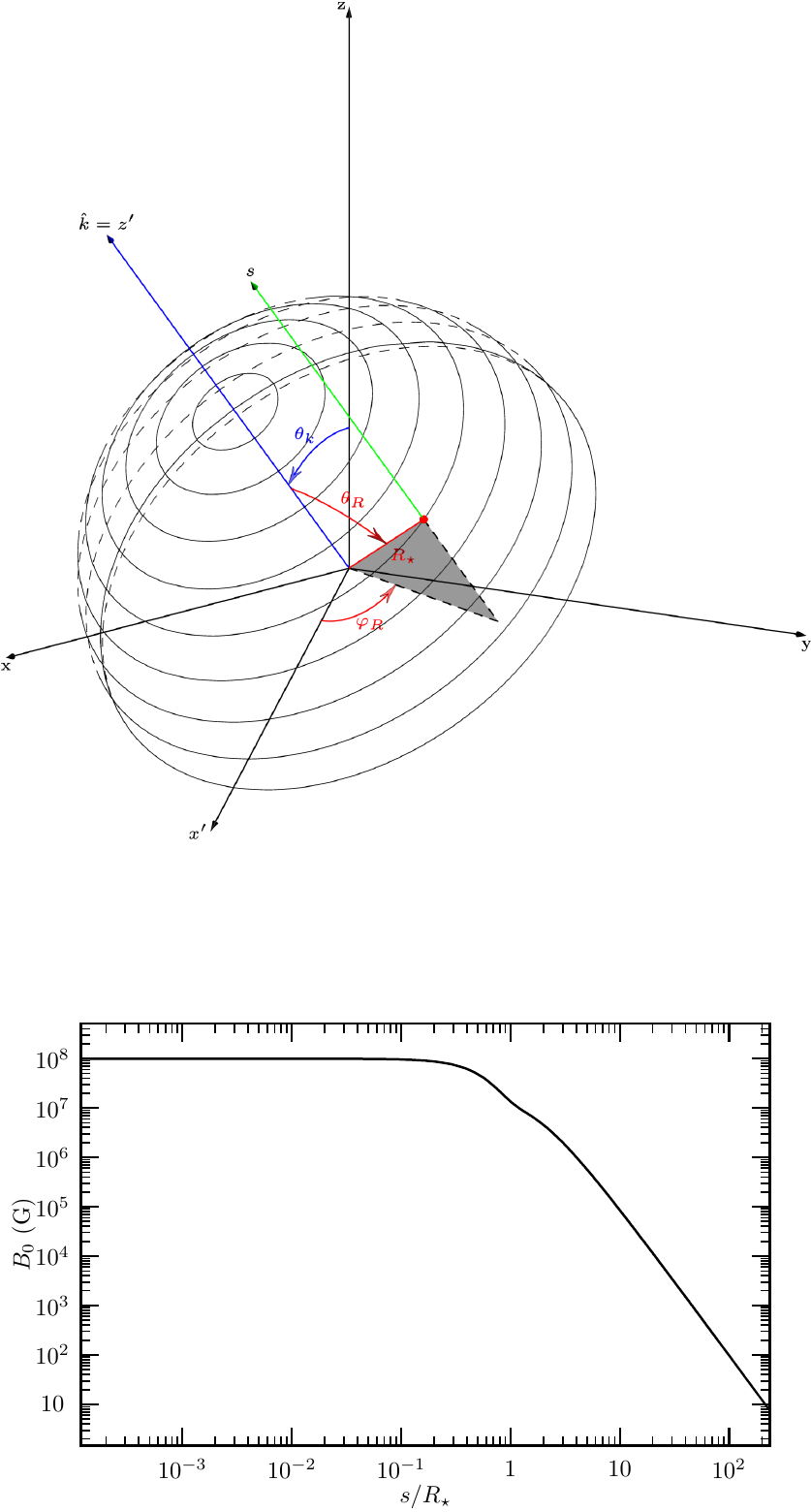}
\caption{This figure illustrates the coordinate system used to obtain the photon-axion 
mode evolution along a given LOS (top), and the decline of the magnetic field strength with distance 
$s$ from the surface (bottom). Here the LOS vector is represented by $s$ that is tilted 
at an angle $\theta_k$ to the rotation axis, and is in parallel to $z'$ which is not to be 
confused with $\hat{z}'$ in Fig. \ref{fig:coord}. Several different 
points on the star's surface with spherical coordinates ($R_\star,\theta_R,\phi_R$) are 
chosen and then averaged to determine the final polarization of the photon leaving the 
magnetosphere (color online).}
\label{fig:los}
\end{figure}
 The Zeeman tomography analysis of mWD PG 1015+014 indicates that the line of sight (LOS) 
is inclined at an angle $\theta_k=23^\circ$ to the rotational axis of the star. 
Following \citep{DupaysRoncadelli2006}, we modify the matrix Eq.(\ref{eq:modeevol}) to 
obtain the mode evolution of the photon-axion system in a coordinate system oriented 
along the LOS (see Fig. \ref{fig:los}). Again, we assume the ansatz $a\propto\exp{(ik's-i\omega t)}$
\begin{equation}
i\frac{d}{ds}\left(\begin{array}{c}
a \\
E_{x'} \\
E_{y'} \\
\end{array}\right) = \left(\begin{array}{ccc}
\Delta_a-k' & \Delta_{Mx'} & \Delta_{My'} \\
\Delta_{Mx'} & i\chi_{11}'-k' & i\chi_{12}' \\
\Delta_{My'} & i\chi_{21}' & i\chi_{22}'-k'\end{array}\right)
\left(\begin{array}{c}
a \\
E_{x'} \\
E_{y'} \\
\end{array}\right) 
\label{eq:newmodeevol}
\end{equation}
where $\Delta_a=m_a^2/2\omega$, $\Delta_{Mx'}=-g_{a\gamma\gamma}B_x/2$, 
$\Delta_{My'}=-g_{a\gamma\gamma}B_y/2$.   
Notice that Eq.(\ref{eq:modeevol}) applies to a system for which the LOS vector coincides with the rotational 
axis of the star. For a different LOS vector, such as shown in Fig. \ref{fig:los}, we perform a rotation of 
the plasma dielectric tensor around the $\mathbf{\hat{y}}$-axis by an angle $\theta_k$, 
$\matlabel{\tilde{\varepsilon}}' = R_y^T(\theta_k)\matlabel{\tilde{\varepsilon}}R_y(\theta_k)$ where $R_y$ is given in 
Eq.(\ref{eq:Ry}). 

The total degree of polarization can be found by integrating Eq.(\ref{eq:newmodeevol}) from a given point on the surface outwards 
to a distance beyond which the amplitude of photon-axion oscillations and plasma effects become 
negligible, and then by averaging the Stokes parameters \citep{RybickiLightman2004} over 
the whole observable hemisphere. 
\begin{eqnarray}
I & = & \|E_{x'}\|^2 + \|E_{y'}\|^2 \\
Q & = & \|E_{x'}\|^2 - \|E_{y'}\|^2 \\
U & = & E_{x'} E_{y'}^\ast + E_{y'} E_{x'}^\ast \\
V & = & -i (E_{x'} E_{y'}^\ast - E_{y'} E_{x'}^\ast)
\end{eqnarray}
The mode amplitudes
are in general complex, and in the above set of equations $\ast$ gives the
complex conjugate. We sample the Stokes vector 
from different points on the surface, with coordinates ($R_\star,\theta_R,\phi_R$), that are spread around the LOS vector 
$\mathbf{\hat{k}}$ with $\Delta\phi_R=30^\circ$ and $\Delta\theta_R=10^\circ$, where $10^\circ \leq \theta_R 
\leq 80^\circ$ and $0^\circ \leq \phi_R \leq 330^\circ$. Because the sampling in the azimuthal 
angle is sparse for larger polar angles, we take a weighted average, as shown below for one of the 
Stokes parameters, to determine the average degree of polarization of the whole hemisphere
\begin{equation}
\langle I\rangle = \frac{\sum_{\theta_R,\phi_R}I(\theta_R,\phi_R)\sin\theta_R}{\sum_{\theta_R}\sin\theta_R}
\end{equation}
\section{Results}
In the following, we look at how an unpolarized photon emitted from the photosphere 
of a mWD gets polarized as it traverses through the magnetosphere. Photon-axion 
interaction and the intervening plasma make the medium birefringent, consequently, altering 
the state of polarization of the unpolarized photon. We obtain the degree of polarization from 
the averaged Stokes parameters
\begin{eqnarray}
P_L & = & \frac{\sqrt{\langle Q\rangle^2 + \langle U\rangle^2}}{\langle I\rangle} \\
P_C & = & \frac{\langle V\rangle}{\langle I\rangle} 
\end{eqnarray}
where $P_L$ and $P_C$ represent linear and circular polarization. 
\begin{figure}
\includegraphics[width=0.48\textwidth]{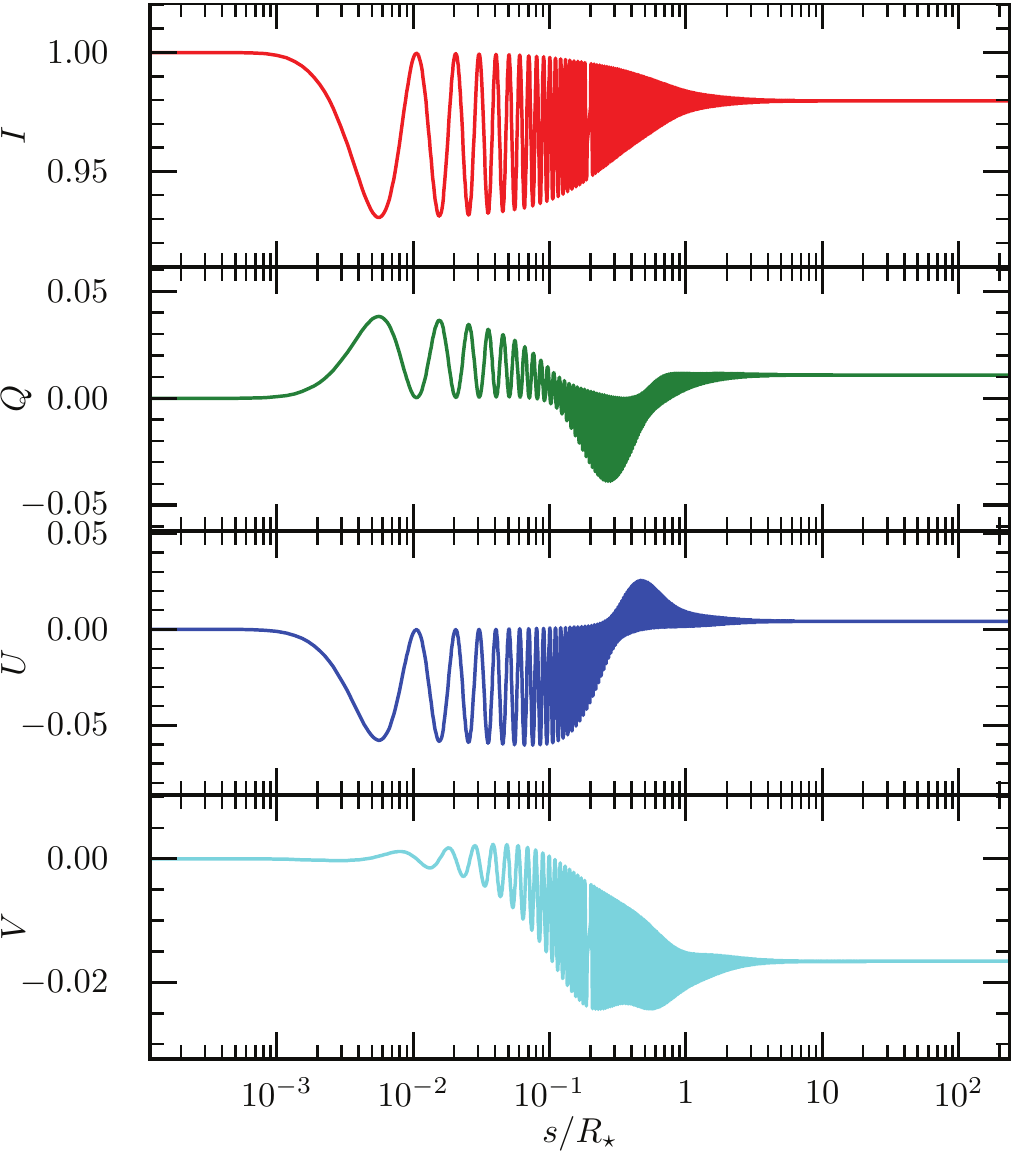}
\caption{Polarization evolution of an unpolarized photon along a given LOS as it starts at 
the photosphere and propagates through the magnetosphere. In this case, $E_\gamma = 3$ 
eV, $m_a = 10^{-5}$ eV, $g_{a\gamma\gamma} = 10^{-9}\mbox{ GeV}^{-1}$. The magnetic field 
geometry assumed is that of mWD PG 1015+014 (color online).}
\label{fig:stokes}
\end{figure}

In Fig. \ref{fig:stokes}, we present the evolution of the Stokes vector with distance $s$ 
from the surface of the star for the case of radiation with $E_\gamma = 3$ eV, and axion parameters 
$m_a=10^{-5}\text{ eV}, g_{a\gamma\gamma}=10^{-9}\text{ GeV}^{-1}$. The oscillations in the solution arise due to the 
mixing of the axion and photon eigenstates, an effect analogous to neutrino oscillations due to the MSW 
effect \citep{MikheevSmirnov1986,Wolfenstein1978}. However, notice that the interaction is non-resonant because a 50\% drop in intensity would be 
observed if the axion and photon modes were to achieve maximal mixing and undergo level crossing. Eventually, 
as the photons travel farther away from the surface, the decline in the magnetic field strength reduces the probability 
of conversion, hence the diminishing of intensity variation. We find that the change in 
polarization is primarily brought about by the axion interaction with the photon. In the event this 
interaction is made negligible, no significant polarization or change in intensity of the emergent radiation 
is found. The origin of circular polarization in mWDs, as alluded 
to earlier, is understood in terms of the difference in opacities for the two modes of radiation, making 
the plasma dichroic, in the presence of a magnetic field. Linear polarization, on the other hand, was 
explained by the cyclotron radiation that emanates from the tenuous corona composed of an ionized plasma. 
In this study, since the treatment of radiative transfer effects is very simplistic only an upper limit can be
placed on how strongly the axion couples to photons, as shown in the next section.
\subsection{Constraints on $g_{a\gamma\gamma}$}
\begin{figure}
\includegraphics[width=0.48\textwidth]{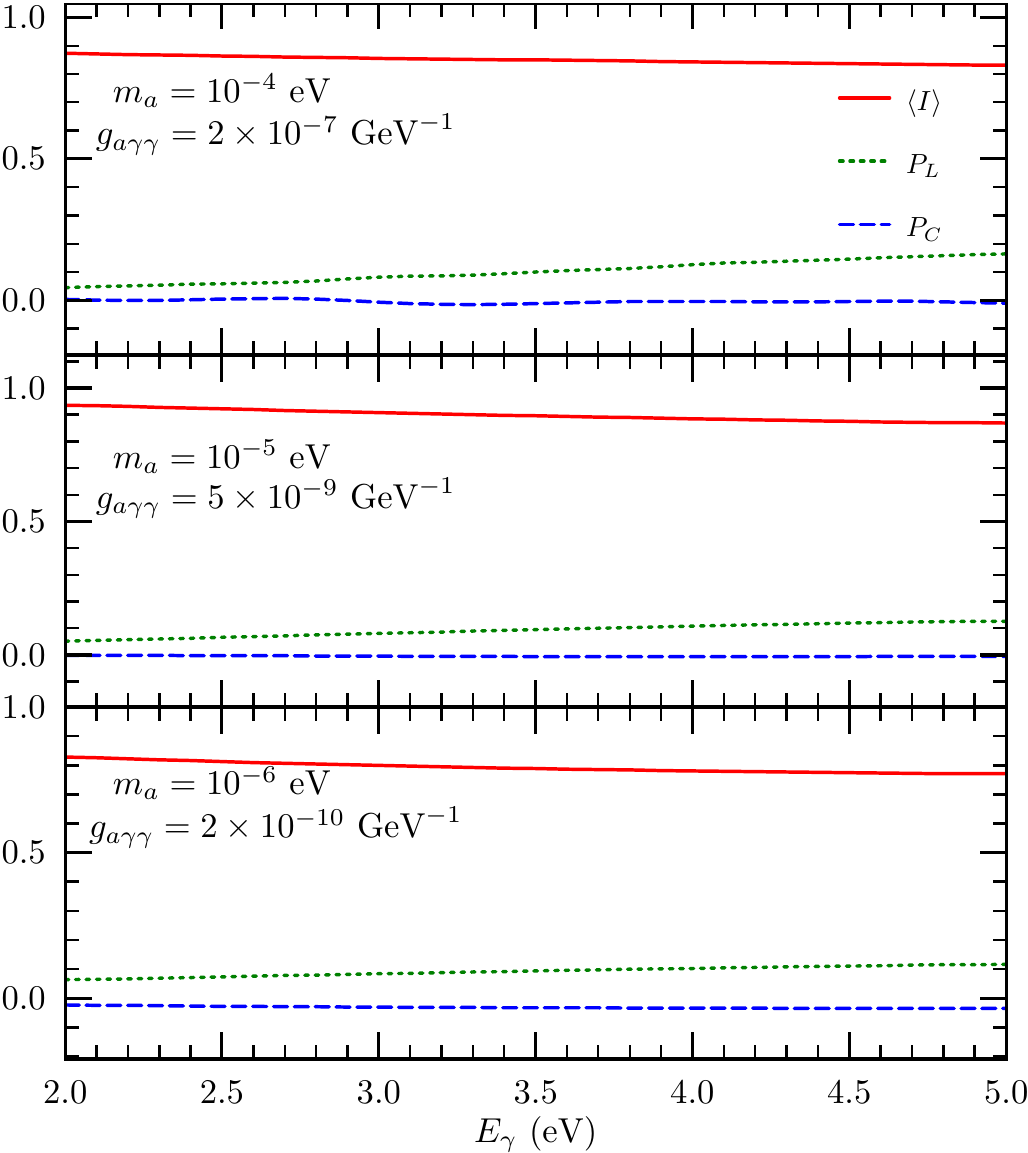}
\caption{The final state of polarization of radiation after traversal from the WD's magnetosphere for different 
$m_a$ and $g_{a\gamma\gamma}$. Here $\langle I \rangle$ is the average Stokes intensity, and $P_L$ and $P_C$ are 
the degrees of linear and circular polarizations. These results apply to the case of mWD PG 1015+014 (color online).}
\label{fig:composite}
\end{figure}
Axion production in the mWD magnetosphere can enhance the degree of linear polarization of the 
observed optical radiation. The goal here is to not determine 
the precise value of the photon-axion coupling strength but only constrain it from above. To this 
end, we look at the amount of linear polarization that is produced for a given $m_a$ and $g_{a\gamma\gamma}$.
The underlying assumption here is that all of the observed linear polarization is generated due to 
photon-axion interaction, and not by the plasma, which effectively yields the absolute upper limit on 
$g_{a\gamma\gamma}$. In Fig. \ref{fig:composite}, we plot the emergent intensity and state of polarization 
for different axion masses and for photons in the UV - optical waveband with energies between $2 - 5$ eV. 
The $m_a$ and $g_{a\gamma\gamma}$ in Fig. \ref{fig:composite} were chosen specifically so that $P_L \gtrsim 0.05$ 
for all photon energies. 

\begin{figure}
\includegraphics[width=0.48\textwidth]{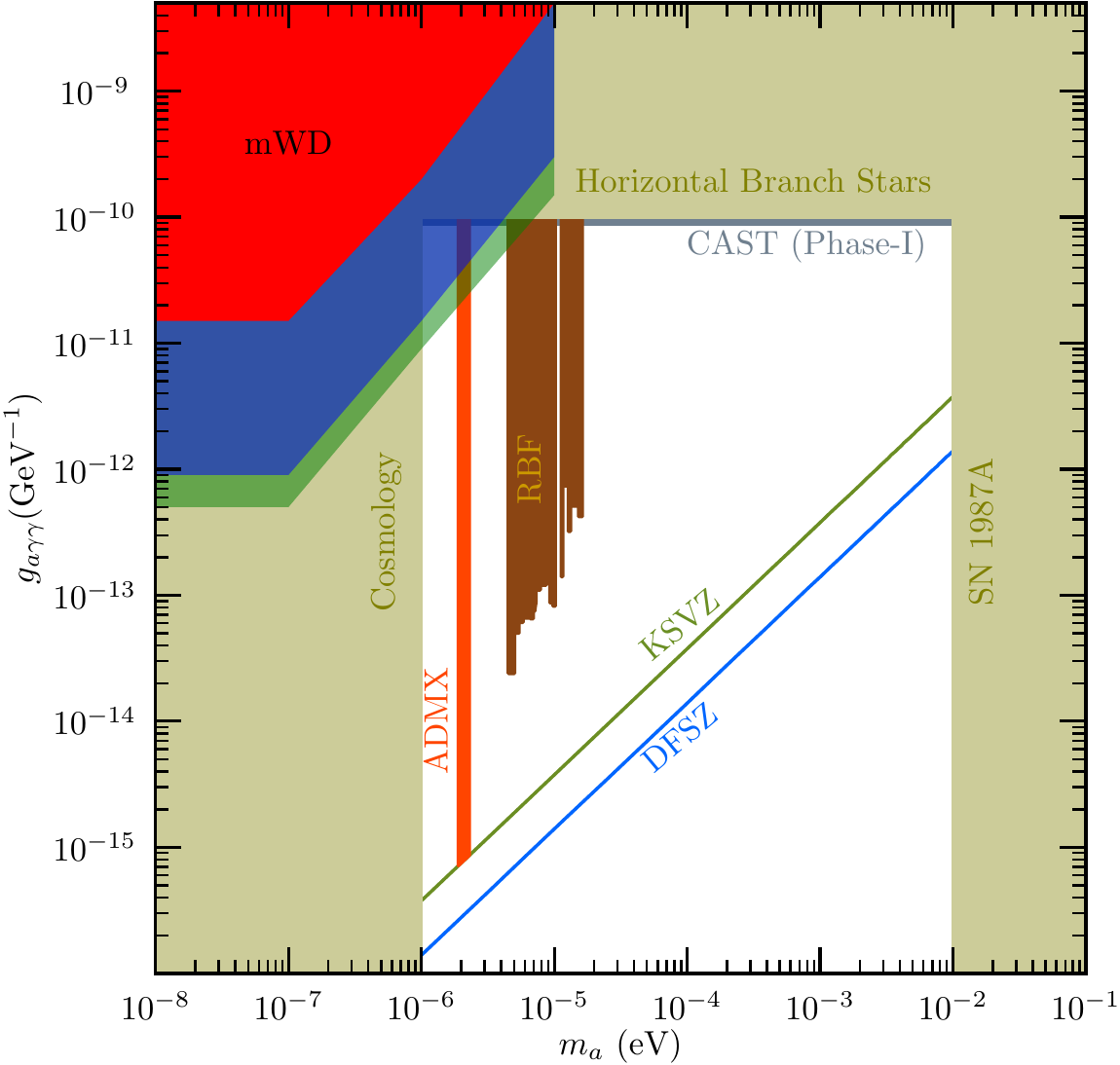}
\caption{Exclusion plot in the $m_a-g_{a\gamma\gamma}$ parameter space. The red region corresponds to 
the case of mWD PG 1015+014 with $B\simeq10^8$ G and limiting linear polarization $P_L\sim5\%$. The 
blue and green regions correspond to the case with $B\simeq10^9$ G, but with $P_L\sim5\%$ and 
$P_L\sim1\%$ respectively. The mass of the 
axion is constrained to $10^{-6} \lesssim m_a \lesssim 10^{-2}$ eV from cosmology and SN 1987A measurments.
The photon-axion coupling constant is capped from above with $g_{a\gamma\gamma} < 10^{-10}
\text{ GeV}^{-1}$ by the number of horizontal branch stars in globular clusters. KSVZ 
\citep{Kim1979,ShifmanVainshteinZakharov1980} and DFSZ \citep{DineFischlerSrednicki1981,Zhitnitskii1980} 
are two different theoretical models that predict how $g_{a\gamma\gamma}$ scales with $m_a$. 
Other exclusion regions are from lab experiments by the CAST experiment \citep{Andriamonjeetal2007}, 
ADMX group \citep{Asztalosetal2004,Asztalosetal2010} and by Rochester-Brookhaven-Fermilab collaboration 
\citep{Panfilisetal1987,Wuenschetal1989} (color online).} 
\label{fig:exclusion}
\end{figure}

In Fig. \ref{fig:exclusion}, we use the same parameters to draw an exclusion region in the $m_a-g_{a\gamma\gamma}$ 
parameter space, along with regions excluded by lab experiments and astrophysical considerations. The shaded region 
in red excludes all $m_a$ - $g_{a\gamma\gamma}$ values for the case of mWD PG 1015+014, that is for a typical 
surface field strength $B\simeq10^8$ G and degree of linear polarization $P_L\sim 5\%$. We find that for 
the range of masses that are of relevance, in particular, to the axion models, the constraints on 
$g_{a\gamma\gamma}$ from this study are superseded by that from horizontal branch (HB) stars. Still, the 
limiting linear polarization criterion used in this study is able to probe smaller $g_{a\gamma\gamma}$ 
values in comparison to works that only look at radiation dimming (for e.g. see \citep{Jimenezetal2011}).

The constraints can be further improved by looking at mWDs with higher magnetic field strengths. The highest 
field strength that has ever been discovered in a mWD is $B\simeq1000$ MG in two such objects namely, 
PG 1031+234 and SDSS J234605+385337 \citep{Jordan2009}. Both objects show linear polarization as low as 
$\sim1\%$ for some rotational phases \citep{PiirolaReiz1992,Vanlandinghametal2005}. Based on these two 
facts and assuming that the magnetic field geometry of these two mWDs is at least as complex as that found 
in PG 1015+014, we produce two exclusion regions shown in Fig. \ref{fig:exclusion} with colors blue 
and green. The former corresponds to a surface field strength $B=1000$ MG with the same level of linear polarization 
as before, and the latter studies the case with $P_L\simeq1\%$. For these two cases, we have only looked at 
$m_a \leq 10^{-5}$ eV since higher mass values don't constrain $g_{a\gamma\gamma}$ better than limits derived 
from HB stars and CAST (Phase-I). On the other hand, we have extended our treatment to smaller particle masses 
with $m_\phi\leq10^{-6}$ eV where $m_\phi$ should be interpreted as the mass of any light pseudoscalar 
boson that is characteristically very much similar to the axion but isn't a CDM particle.

It is worth mentioning that the change in $g_{a\gamma\gamma}$ is not linear with the change in magnetic 
field strength, as evident from the comparison between the red and blue regions in Fig. \ref{fig:exclusion}.
For higher field strengths one observes higher degree of polarization of the emerging radiation. Naively, one 
would expect the plane of polarization to rotate by an amount that is $\mathcal{O}(g_{a\gamma\gamma}^2 
B^2l^2)$, which is valid strictly in the absence of plasma when the photon and axion are treated as 
massless particles \citep{VanBibberetal1987,RaffeltStodolsky1988}. Therefore, for a fixed degree of 
polarization an increase in $B$ should also decrease $g_{a\gamma\gamma}$ by the same factor, when 
$l$, the length over which the magnetic field remains homogeneous, is kept constant. However, as 
shown by \citep{Heyletal2003} in the case of NSs, an increase in magnetic field strength also 
increases the level of polarization by effectively shifting the polarization-limiting radius, 
the distance beyond which the two polarization modes couple and which depends weakly on 
the magnetic field strength $R_{pl}\propto B^{2/5}$, farther away from the star. The farther the 
polarization-limiting radius, the more coherently the polarization states from different LOSs add, 
yielding a higher degree of polarization.
\section{Discussion}
This study looks at how the production of axions in mWD magnetospheres can alter the 
state of polarization of the observed radiation. We find that unpolarized photons of 
photospheric origin become linearly polarized upon their traversal through the 
inhomogeneous magnetic field of a mWD. We have modeled the magnetospheric plasma, as fully 
ionized pure H with a barometric profile. Since the majority of mWDs are strongly 
circularly polarized and only show a relatively 
small degree of linear polarization, at most $5\%$, we have used this observation to constrain the 
coupling strength $g_{a\gamma\gamma}$ of axions to photons. We find that for the case where 
the plasma component only contributes negligibly to the state of polarization, the 
coupling strength $g_{a\gamma\gamma}$ increases with the mass of the axion $m_a$. The level 
of linear and circular polarization observed in mWDs is sensitive to the properties of the 
magnetospheric plasma. The limits on $g_{a\gamma\gamma}$ can be improved by modelling all 
the radiative transfer effects in WD atmospheres and fitting the model spectra to real observations.

Magnetic fields stronger than that of mWDs exist in NSs. Going back to the argument of how 
astrophysical objects, compared to laboratory experiments, benefit from longer coherence 
lengths (see Section I), in comparing mWDs with NSs, one finds that the latter are $\sim 10^4$ 
times more efficient in converting photons to axions and vice-versa. A number of studies have expounded 
on the subject of propagation of polarized radiation through the NS magnetosphere, where 
they have considered IR/Optical radiation \citep{ShannonHeyl2006}, and thermal X-rays 
\citep{LaiHo2003a,LaiHo2003b} produced at the surface of the NS. Unfortunately, no X-ray 
polarimetry observations have been conducted partly due to the very low flux in X-rays 
from these objects, and also because none of the high energy telescopes are equipped 
with a polarimeter. X-ray polarimetry has been neglected for the last 30 years but it 
is hoped that some of the future space missions \citep{Mulerietal2010}, for example the Gravity and Extreme 
Magnetism Small Explorer (GEMS) \citep{Jahoda2010}, will fill this 
void in X-ray astronomy. In any case, as discussed by \citep{RaffeltStodolsky1988,
LaiHeyl2006,Cheloucheetal2009}, NSs are excellent laboratories for the detection of 
any light, weakly coupled pseudoscalar particle. 
\subsection{Outlook}
The ADMX\footnote{http://www.phys.washington.edu/groups/admx/experiment.html} 
project, that employs a microwave cavity to search for cold dark matter axions, 
will begin its phase II of testing in the year 2012. With the new 
upgrades the ADMX project will be able to exclude $g_{a\gamma\gamma}$ up to the DFSZ line in 
the same mass range as before. Although outside of the range of axion masses probed 
in this study, the modified CAST experiment has been able to exclude axions with 
$g_{a\gamma\gamma} \gtrsim 2.2\times10^{-10}\mbox{ GeV}^{-1}$ for $m_a \lesssim 0.4$ 
eV, becoming the first experiment to ever cross the KSVZ line \citep{Ariketal2009,Cast2011}. 
The currently running CAST experiment in its phase-II will 
be able to exclude axions with $m_a\lesssim 1.15$ eV with unprecedented sensitiviy in 
this mass range. An improved version of the \textit{light shining through wall} (LSW) experiment 
(\citep{VanBibberetal1987}, also see \citep{RedondoRingwald2011} for a recent review on 
such experiments) using Fabry-Perot optical cavities to resonantly enhance photon-axion conversion 
has been proposed \citep{HoogeveenZiegenhagen1991,Muelleretal2009,Sikivieetal2007}. The projected limit in sensitivity to 
$g_{a\gamma\gamma} \gtrsim 2.0\times 10^{-11}\mbox{ GeV}^{-1}$ typically for 
axion masses $m_a \lesssim 10^{-4}$ eV achieved using 12 Tevatron superconducting 
dipoles appears quite promising. Further improvements in experiment design and optimization techniques 
yielding increased sensitivity to even smaller coupling strengths have also been suggested by many workers 
in the field, for example the use of the dipole magnets, each providing a field strength of 5 T, 
from the Hadron Electron Ring Accelerator (HERA) at DESY in 
Hamburg in a 20+20 configuration can potentially exclude 
$g_{a\gamma\gamma} \gtrsim 10^{-11}\text{ GeV}^{-1}$ for $m_a < 10^{-4}$ eV \citep{Ringwald2003,Ariasetal2010}. 
Another proposed line of investigation to search for axion like particles (ALPs) is the use of resonant microwave cavities which are 
much similar in design to the optical LSW experiments discussed above \citep{Hoogeveen1992,Caspersetal2009,JaeckelRingwald2008}. This 
method has already been employed to search for hidden sector photons \citep{Poveyetal2010} and can prove to be a powerful 
tool in the case of axions.

Finally, the simplistic model assumed for the mWD atmosphere only yields an absolute upper bound 
on $g_{a\gamma\gamma}$. A much tighter constraint can be obtained by adopting a more realistic 
atmospheric model and solving the equations of radiative transfer with the photon-axion oscillations 
included. Such an analysis is outside the scope of this study, but it is hoped that the novel 
method discussed in this work will prove to be extremely useful in better constraining the 
properties of any ALP.
\begin{acknowledgements}
   We would like to thank the referee for significantly improving the 
   quality of this work. R.G. acknowledges support by the NSERC CGS-D3 scholarship. 
   The Natural Sciences and Engineering Research Council of Canada,
   Canadian Foundation for Innovation and the British Columbia
   Knowledge Development Fund supported this work. Correspondence and
   requests for materials should be addressed to
   J.S.H. (heyl@phas.ubc.ca). This research has made use of NASA’s
   Astrophysics Data System Bibliographic Services
\end{acknowledgements}
\bibliographystyle{prsty}

\end{document}